\documentclass{PoS}

\title{Contribution to diffuse gamma-ray emission coming from self-confined CRs around their Galactic sources}

\ShortTitle{$\gamma$-ray emission from self-confined CRs}

\author{\speaker{Giovanni Morlino}
	\\
        Gran Sasso Science Institute, viale F. Crispi 7, 67100 L'Aquila, Italy.\\
        E-mail: \email{giovanni.morlino@infn.gssi.it}}

\author{Marta D'Angelo\\
        Gran Sasso Science Institute\\
        E-mail: \email{marta.dangelo@infn.gssi.it}}

\author{Pasquale Blasi\\
        INAF-Osservatorio Astrofisico di Arcetri {\it and} Gran Sasso Science Institute\\
        E-mail: \email{blasi@arceri.astro.it}}

\author{Elena Amato\\
        INAF-Osservatorio Astrofisico di Arcetri \\
        E-mail: \email{amato@arceri.astro.it}}

\abstract{
Recent observations of the diffuse Galactic $\gamma$-ray emission by the Fermi-LAT satellite have shown significant deviations from models which assume the same diffusion properties for cosmic rays (CR) throughout the Galaxy.
We explore the possibility that a fraction of this diffuse Galactic emission could be due to hadronic interactions of CRs self-confined in the region around their sources. In fact, freshly accelerated CRs that diffuse away from the acceleration region can trigger the streaming instability able to amplify magnetic disturbance and to reduce the particle diffusion. When this happen, CRs are trapped in the near source region for a time longer than expected and an extended $\gamma$-ray halo is produces around each source.
Here we calculate the contribution to the diffuse $\gamma$-ray background due to the overlap along lines of sight of several of these extended halos.  We find that if the density of neutrals is low, the halos can account for a substantial fraction of the diffuse emission observed by Fermi-LAT, depending on the orientation of the line of sight with respect to the direction of the galactic center.}

\FullConference{35th International Cosmic Ray Conference\\
         10-20 July, 2017\\
         Bexco, Busan, Korea}

\begin{document}

\section{Introduction}		\label{sec:intro}
The spectrum and morphology of the diffuse Galactic $\gamma$-ray emission carries valuable information on the cosmic ray (CR) spectrum. Recent results obtained by analyzing Fermi-LAT data accumulated over 7 yr of observation show a substantial variation of the CR spectrum as a function of the distance from the Galactic Centre \cite{Acero2016,Yang2016}. In particular the density in the central region of the Galaxy was found to exceed the value measured in the outer Galaxy, even when the larger number of supernova remnants (SNR) in the inner disc is taken into account. Moreover, data suggest a spectral hardening toward the inner Galaxy, with a slope $s\simeq2.4$ for photon energy above 2 GeV, to be compared with the outer Galaxy where $s \simeq 2.7$. Such a findings are a challenge for models that assume: (1) that all SNRs eject the same CR spectrum and (2) that the diffusion coefficient in uniform in the whole Galactic volume. 
While it is difficult to justify a model where the acceleration properties of SNRs depend on the galactocentric distance, it is easy to figure out why the diffusion should change. In fact, the diffusion depends on the amount of local magnetic turbulence which has two main contributions, one coming from the very large scale turbulence (presumably injected by the same SN explosion) that subsequently cascade at smaller wavelength and the second due to self-generated waves produced through CR streaming \cite{Skilling1971}. 
Indeed, in \cite{Recchia2016} it was shown that when self-generated turbulence is taken into account, both the spatial profile of CR density and the spectral slope can easily be accounted for.
In this work we explore a different prospective, always connected to the non-linear propagation of CRs. When freshly accelerated CRs escape from the source towards the interstellar medium, a large CR density gradient develops resulting in a strong amplification of magnetic turbulence. As a consequence CRs result self-confined around the source for a time much longer than that estimated assuming a diffusion equal to the average galactic one. When this happen the associated $\gamma$-ray emission produced by hadronic interactions is also enhanced and an extended $\gamma$-ray halo will surround the source for a time longer than the typical acceleration time. While the emission from a single halo is too faint to be detectable with current telescopes, the sum over all halos could give a non negligible contribution to the diffuse $\gamma$-ray background detected by Fermi-LAT.

Some investigations of the non-linear problem in which CRs affect their own diffusion have been previously presented by \cite{Ptuskin2008, Malkov2013} using self-similar solutions. In this work we follow the approach of \cite{Dangelo2016} (also similar to the work by \cite{Nava2016}) where the time dependent transport equation  is solved numerically. This is essential to correctly account for both the self-generation and the damping processes.

\section{CR propagation in the vicinity of the source}		\label{sec:conf}
In order to calculate the density of CRs around their sources, we follow the same method developed in \cite{Dangelo2016} where the propagation of particles after the release from the source is coupled to the evolution of the magnetic perturbations responsible for the scattering process.
When particles escape from the acceleration region, they start diffusing around the SNR mainly along the direction of the large scale magnetic field $\mathbf{B}_0 = B_0 \mathbf{z}$. Inside the Galactic disc the magnetic field has a typical coherence scale $L_c \approx 100$ pc~\cite{Beck2016}, hence in a region of size $L_c$ around the SNR (located at $z=0$) we can approximate the transport of particle to occur inside a cylindric flux tube. This 1D geometry breaks down for distances larger than $L_c$, when magnetic field lines tangle, losing their coherent structure. Inside the flux tube the 1D time dependent transport equation for CRs can be written as
\begin{equation}
\frac{\partial f}{\partial t} + u \frac{\partial f}{\partial z} - \frac{\partial }{\partial z} \left[ D(p,z,t) \frac{\partial f}{\partial z} \right] - \frac{du}{dz} \frac{p}{3}\frac{\partial f}{\partial p}= 0 \, .
\label{eq:transp}
\end{equation}
The plasma around the SNR is assumed at rest, so that the advection is determined solely by the bulk motion of  Alfv\'en turbulence which moves at $u \equiv v_A = B_0/\sqrt{4 \pi n_d m_p}$, directed away from the SNR because the self-generated Alfv\'en waves move down the CR gradient.

Eq.~(\ref{eq:transp}) does not contains any source term because we assume that the source lasts for a short time compared to all times of interest. The injection of particles is instead mimicked using an appropriate initial condition at $t=0$, i.e.  
\begin{equation}
f(p,z,t=0) = q_0(p) \exp\left[-\left ( z/z_0 \right)^2\right] + f_g(p) \, .
\label{eq:ic}
\end{equation}
where the Gaussian component represents the CR cloud released by the SNR, while $f_g$ is the pre-existing average Galactic CR spectrum, assumed to be equal everywhere inside and outside the flux tube, and given by the CR proton resulting from the most recent AMS-02 measurements~\cite{AMS02proton}:
\begin{equation}
f_g(p) = 6.8 \times 10^{22} \left(\frac{p}{45 p_0}\right)^{-4.85} \left[1 + \left ( \frac{p}{336 p_0} \right )^{5.54}\right]^{0.024} \mathrm{\left(\frac{erg}{c}\right)^{-3}} \mathrm{cm^{-3}} \, ,
\label{eq:fg}
\end{equation}
where $p_0 = m_p c$.
The solution of Eq.(\ref{eq:transp}) requires two more boundary conditions. One is obtained integrating Eq.(\ref{eq:transp}) around $z=0$ while for the second we impose that the distribution function reduces to the Galactic one at $z=L_c$.
We chosen the size of the gaussian $z_0 = 1$ pc, in order to be $\ll L_c$, but we also checked that our results do not change significantly if $z_0$ changes within a factor of few. 
The function $q_0(p) = A (p/p_0)^{-\alpha}$, with $\alpha= 4$, represents the injection spectrum released by the SNR shock. The normalization constant $A$ is chosen imposing that a fraction $\xi_{CR}=20\%$ of the total kinetic energy of the SNR is released in the form of accelerated particles, i.e. $A = \frac{\xi_{CR} E_{SN}}{\pi R_{SNR}^2 \mathcal{I}}$, where we assumed $E_{SN} = 10^{51}$ erg and $R_{SNR} \approx 20$ pc, which corresponds to the typical value of a SNR radius during the Sedov-Taylor phase.

The particle propagation is determined by the spatial- and time-dependent diffusion coefficient, $D(p,z,t)$ which, in turn, is determined by the turbulence excited by the particles.
Under the assumption that the strength of magnetic turbulence remains much smaller than the large scale field, i.e. $\delta B/B_0 \ll 1$, one can write~\cite{Bell1978}:
$
D(p,z,t) = \frac{1}{3} v(p) r_L(p)  {\mathcal{F}(k, z, t)}^{-1} | _{k = 1/r_L(p)}
$
where $\mathcal{F}(k, z, t)$ is the turbulent magnetic energy density per unit logarithmic bandwidth of waves with wavenumber $k$, normalized to the background magnetic energy density $B_0^2/(8\pi)$. 
Before energetic particles start affecting the environment, the diffusion coefficient is assumed to be the same as the Galactic average one, $D_g$. Assuming a Kolmogorov's turbulence spectrum, we adopt for $D_g$ the analytical expression derived by ~\cite{Ptuskin2009} as a fit to GALPROP results within a leaky box model, i.e. $D_g(E) = 3.6 \times 10^{28} E_{\mathrm{GeV}}^{1/3} \, \mathrm{cm^2 s^{-1}}$. 

The 1D approximation is only valid to describe the propagation of particles up to a distance $|z|<L_c$, and the obvious requirement is that the particle mean free path be less than $L_c$. In mathematical terms this condition is written as $3 D_g(E)/c \ll L_c$, and is satisfied for particle energies up to $10^6 \, \mathrm{GeV}$. In the following we restrict our analysis to CRs with energy in the range $ 10 \, \mathrm{GeV} \leq E \leq 10^4 \, \mathrm{GeV}$.

For $t > 0$ the evolution of the self-generated turbulence $\mathcal{F}(k, z, t)$ is regulated by the balance between growth and damping of waves and is described by the following wave equation
\begin{equation}
\frac{\partial \mathcal{F}}{\partial t} + u \frac{\partial \mathcal{F}}{\partial z} = (\Gamma_{CR} - \Gamma_D) \mathcal{F}(k, z, t) \, ,
\label{eq:calF}
\end{equation}
which we solve in the flux tube $0 \leq z \leq L_c$ where $u=v_A$. The evolution of $\cal F$ is determined by the competition between wave excitation by CRs at a rate $\Gamma_{CR}$ and wave damping, which occurs at a rate $\Gamma_D$. The growth rate due to resonant amplification depends on the CR distribution function $f(p,z,t)$ as~\cite{Skilling1971}
\begin{equation}
\Gamma_{CR} = \frac{16 \pi^2}{3} \frac{v_A}{\mathcal{F} B_0^2} \left |p^4 v(p) \frac{\partial f}{\partial z}\right|_{p = qB_0/(kc)} \, .
\label{eq:GammaCR}
\end{equation}
The main damping mechanisms that can be at work in the situation we consider are: non linear damping~\cite{Ptuskin2003}, ion-neutral damping~\cite{Kulsrud1969} and damping due to pre-existing magnetic turbulence~\cite{FarmerGoldreich2004}. 
Non-linear Damping (NLD hereafter) is due to wave-wave interactions and well describes the evolution of a Dirac-delta function in $k$-space towards a Kolmogorov spectrum of waves. Its rate can be written as~\cite{Ptuskin2003}
$\Gamma_{\rm NLD}(k,z,t) = (2 c_k)^{-3/2} k \, v_A \sqrt{\mathcal{F}(k,z,t)} $,
where $c_k \approx 3.6$. 
Ion neutral damping (IND) is a mechanism that only operates in partially ionized plasmas and is due to the charge-exchange process which couples ions and neutrals causing a friction on the wave propagation. The general expression for the IND damping rate is provided by \cite{Kulsrud1969}.
%
%
%
%
Finally, the damping proposed by Farmer \& Goldreich \cite{FarmerGoldreich2004} is also due to wave-wave coupling, like the NLD, but with a pre-existing MHD turbulence. In our case such a pre-existing turbulence is provided by the Kolmogorov spectrum assumed to pervade the galactic ISM. The damping rate associated with this process reads $\Gamma_{\rm FG} = k v_A \left(k L_{\rm MHD} \right)^{-1/2}$,  where $L_{\rm MHD}$ is the characteristic scale of the background MHD turbulence assumed to be equal to $L_c$.

Now, Eqs.~(\ref{eq:transp}) and (\ref{eq:calF}) must be solved simultaneously. The numerical procedure adopted is based on a finite difference method for the discretization of partial derivatives (both in space and time) and a backward integration in time for Eq.~(\ref{eq:transp}). 

\begin{figure}[t]
\includegraphics[width=0.49\columnwidth]{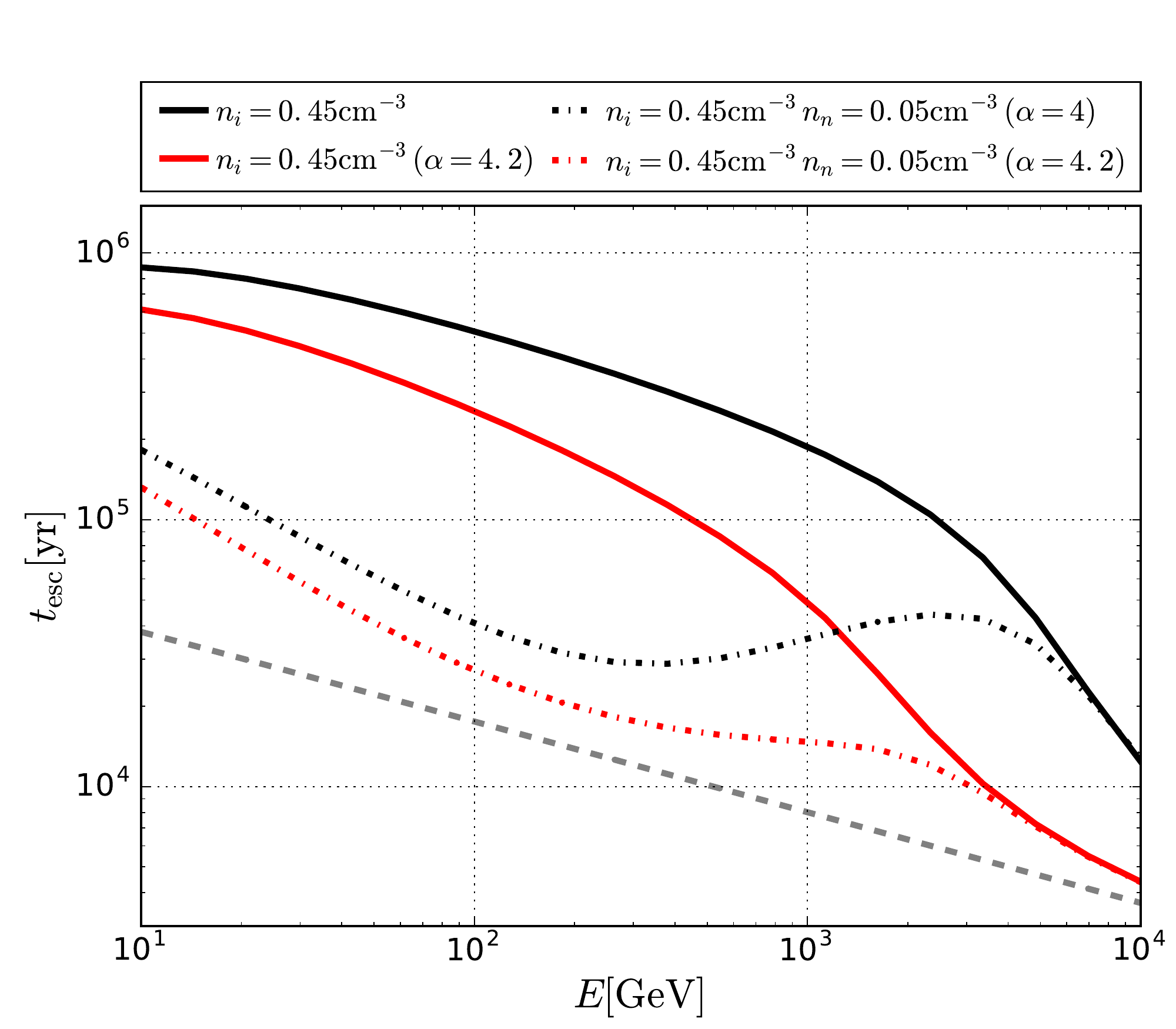}
\includegraphics[width=0.49\columnwidth]{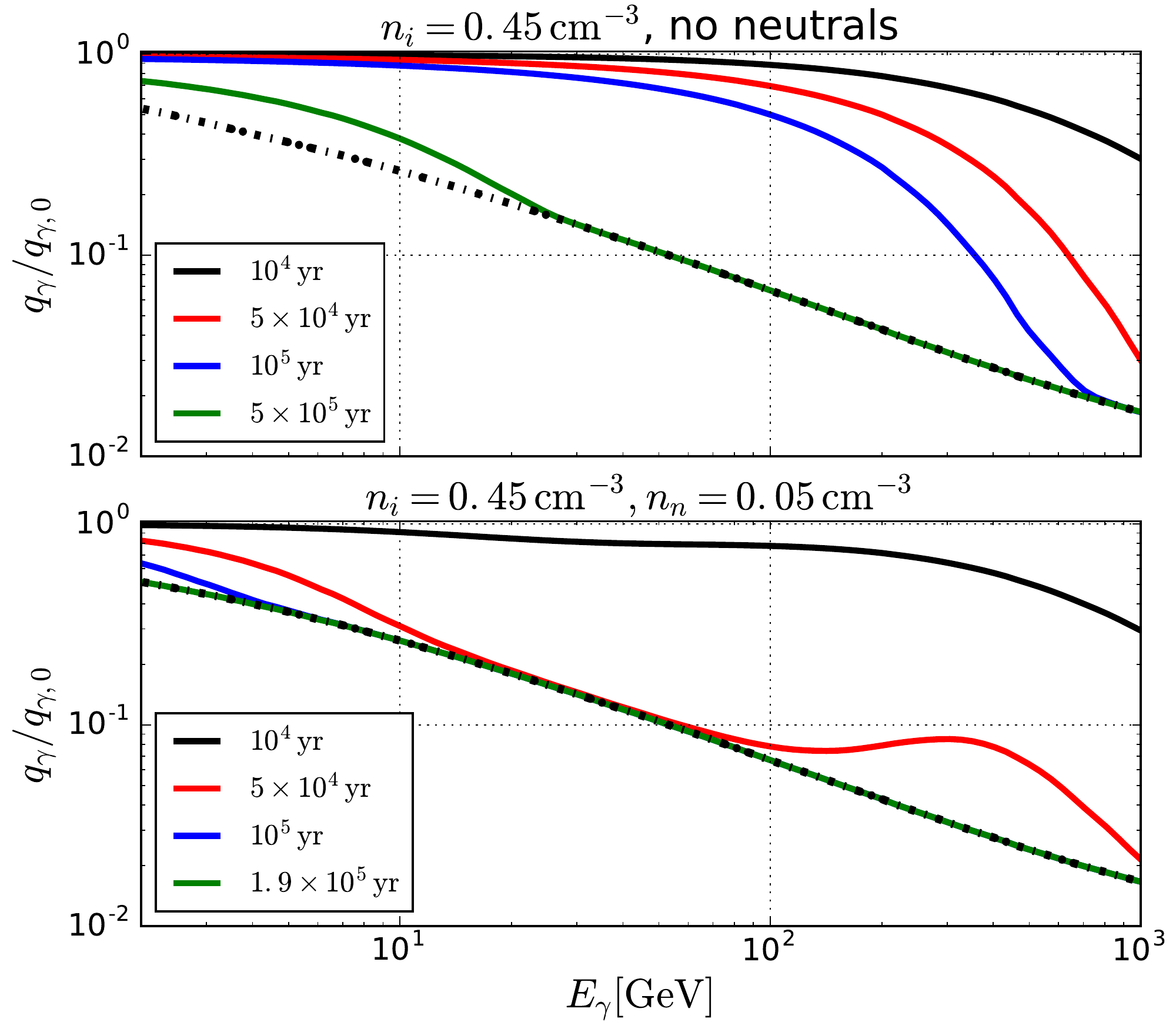}
\caption{{\it Left panel}: CR escape time $t_{\rm esc}$ as a function of particle energy for two different types of ISM: case (1) no neutrals and $n_i = 0.45\,  \mathrm{cm^{-3}}$ (solid lines); case (2) $n_{n}=0.05 \, \mathrm{cm^{-3}}$ and $n_i = 0.45 \, \mathrm{cm^{-3}}$ (dot-dashed lines). Black (red) lines are calculated for injection spectrum with a slope $\alpha = 4.0 (4.2)$.
The grey dashed line is the standard diffusion time calculated with the average Galactic diffusion coefficient, i.e. $\tau_d = L_c^2/D_g = 8.3 \times 10^4 (E_{\mathrm{GeV}})^{-1/3} \, \mathrm{yr}$. 
{\it Right panel}: Normalized $\gamma$-ray emission $q_{\gamma}(E_{\gamma},t)/q_{\gamma}(E_{\gamma},t = 0)$ (integrated over the flux tube volume) for case (1) (top) and case (2) (bottom) at four different times as written in the labels.
The dot-dashed black line represents the $\gamma$-ray emission from the Galactic CR pool integrated over the same volume, i.e. $f = f_g$.}
\label{fig:tesc}
\end{figure}

Given that neutral friction is likely the most effective wave damping mechanism, depending on the ionization fraction of the medium surrounding the source, we have tried to assess its importance considering two different types of ISM (see \cite{Ferriere2001}). Case (1): fully ionized medium with ion density $n_i=0.45~\rm cm^{-3}$; case (2): partially ionized medium with neutral density $n_{n}=0.05~\rm cm^{-3}$ and ion density $n_{i}=0.45~\rm cm^{-3}$. 
The difference can be appreciated from the left panel of Fig.~\ref{fig:tesc} where we plot the confinement time of particles inside the flux tube, $t_{\rm esc}$. Such a quantity is defined in analogy with the case of spatially constant diffusion.  In the case where a burst of particles is injected a $t=0$ with a preassigned diffusion coefficient equal to $D_g(p)$, we find that about $89\%$ of the total injected particles leave the box within the classical diffusion time $\tau_d=L_c^2/D_g$. We use the same criterion in the case of non-linear diffusion defining $t_{\rm esc}$ as the time when $89\%$ of the injected particles leave the box. One can see that when neutrals are absent (solid black line) $t_{\rm esc}$ is a factor $\sim 20$ larger than $\tau_d$ (gray-dashed line) up to energies around 1 TeV above which the streaming instability is less efficient and the confinement time decreases faster than $E^{-1/3}$. 
On the contrary, when neutrals are present, the IND efficiently damps the turbulence and $t_{\rm esc}$ decreases accordingly but it still remains larger than $\tau_d$. In Fig.~\ref{fig:tesc} we also report the confinement time resulting when the slope of injected spectrum is $\alpha= 4.2$ (red lines), keeping all other parameters fixed: at $E \sim 10$ GeV,  $t_{\rm esc}$ decreases by less than a factor of 2 with respect to the case $\alpha= 4$, while at higher energies the discrepancy increases, as expected.

\section{Comparison with Galactic diffuse emission}		\label{sec:galactic}
The self-confinement process described above can give rise to the formation of an extended halo of gamma radiation produced by $pp$ collisions and subsequent decay of $\pi^0 \to \gamma\gamma$ (see \cite{Aharonian2000} and references therein).
To estimate the contribution of these extended halos to the diffuse galactic $\gamma$-ray background, we start by evaluating the emissivity associated to a single SNR halo, then, assuming some distribution of SNRs in the Galaxy, we finally proceed to summing up all the contributions from SNR halos within given regions of space for direct comparison with available data.

Using the numerical solutions of the coupled Eqs.~(\ref{eq:transp}) and~(\ref{eq:calF}), we calculate the time-dependent CR spectrum associated with a single SNR, integrating over a cylindrical flux tube of radius $R_{SN}$ and length $2 L_c$ i.e. $J_p(E, t) = 2 \pi R_{SN}^2 \int_0^{L_c} dz \, f(E,z,t)$. The volume-integrated gamma-ray emission $q_{\gamma}(E_{\gamma},t)$ is then computed using the $\delta$-function approximation detailed in the works by \cite{Aharonian2000} and \cite{Kelner2006}. 
In the right panel of Fig.~\ref{fig:tesc} we show the volume-integrated gamma-ray emission, $q_{\gamma}(E_{\gamma},t)$, at different times during the SNR evolution, between $10^4$ and $few \times 10^5$ yr and for the two different ambient conditions case (1) and (2).
All curves are normalized to their value at time $t=0$, with the notation $q_{\gamma}(E_{\gamma},t = 0) \equiv q_{\gamma,0}$, which represent the peak value. 
In both cases, at time $t \lesssim 10^4 \, \mathrm{yr}$ CRs with $E \lesssim 10^3$ GeV are all confined within the box, which implies that for $E_{\gamma} \lesssim 10^2 \, \mathrm{GeV}$ one has $q_{\gamma} \simeq q_{\gamma,0}$. In the case (1) the confinement is effective up to $t \simeq 10^5$ yr while for times $\gtrsim 5 \times 10^5$ yr the spectrum reduces to the average Galactic one.
In the case (2), because of the presence of neutrals, the escape of particles is much faster and the flux tube around the source is depleted already at $t \gtrsim 5 \times 10^4$.

\begin{figure}[t]
\centering
\includegraphics[width=0.49\columnwidth]{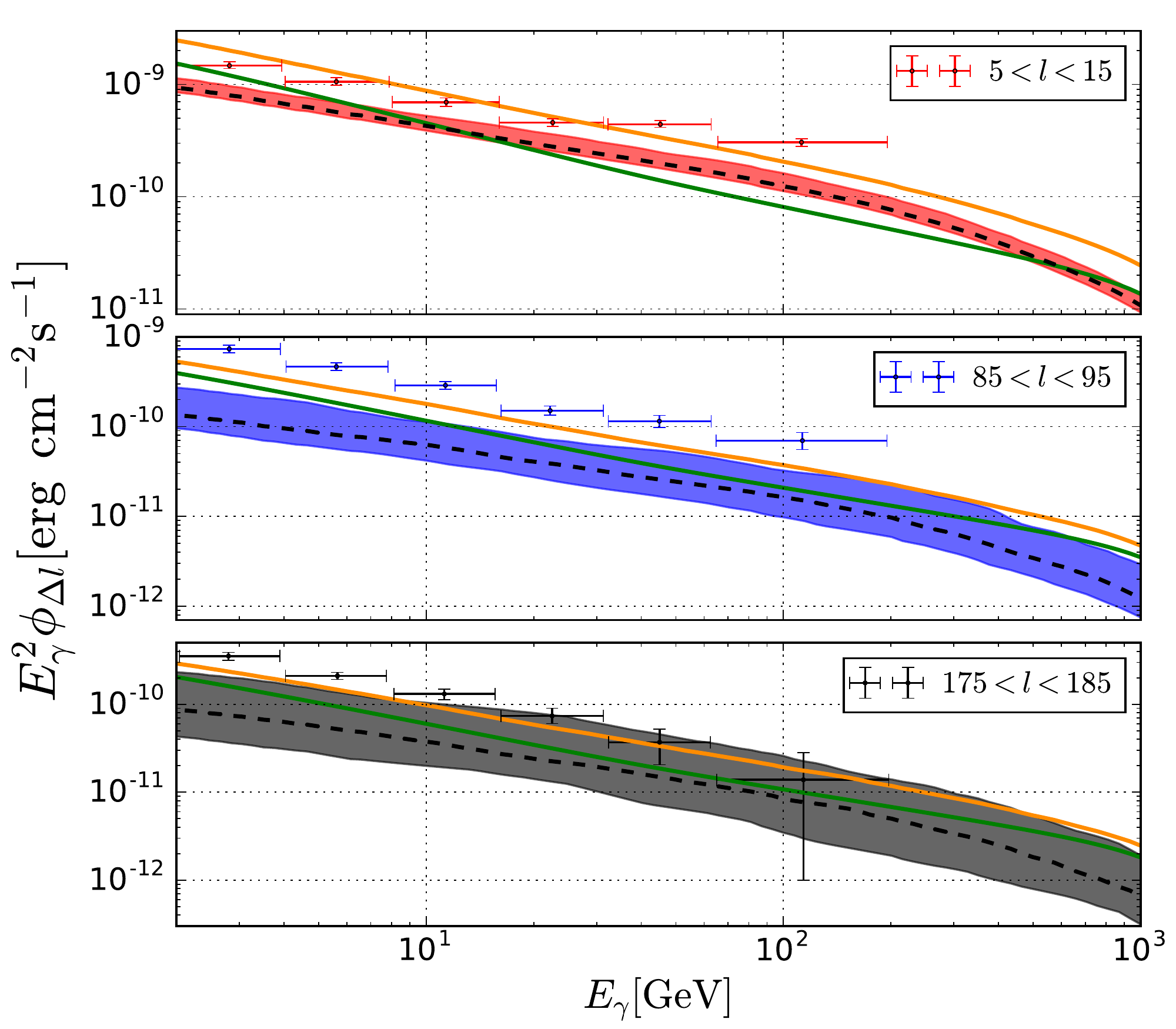} 
\includegraphics[width=0.49\columnwidth]{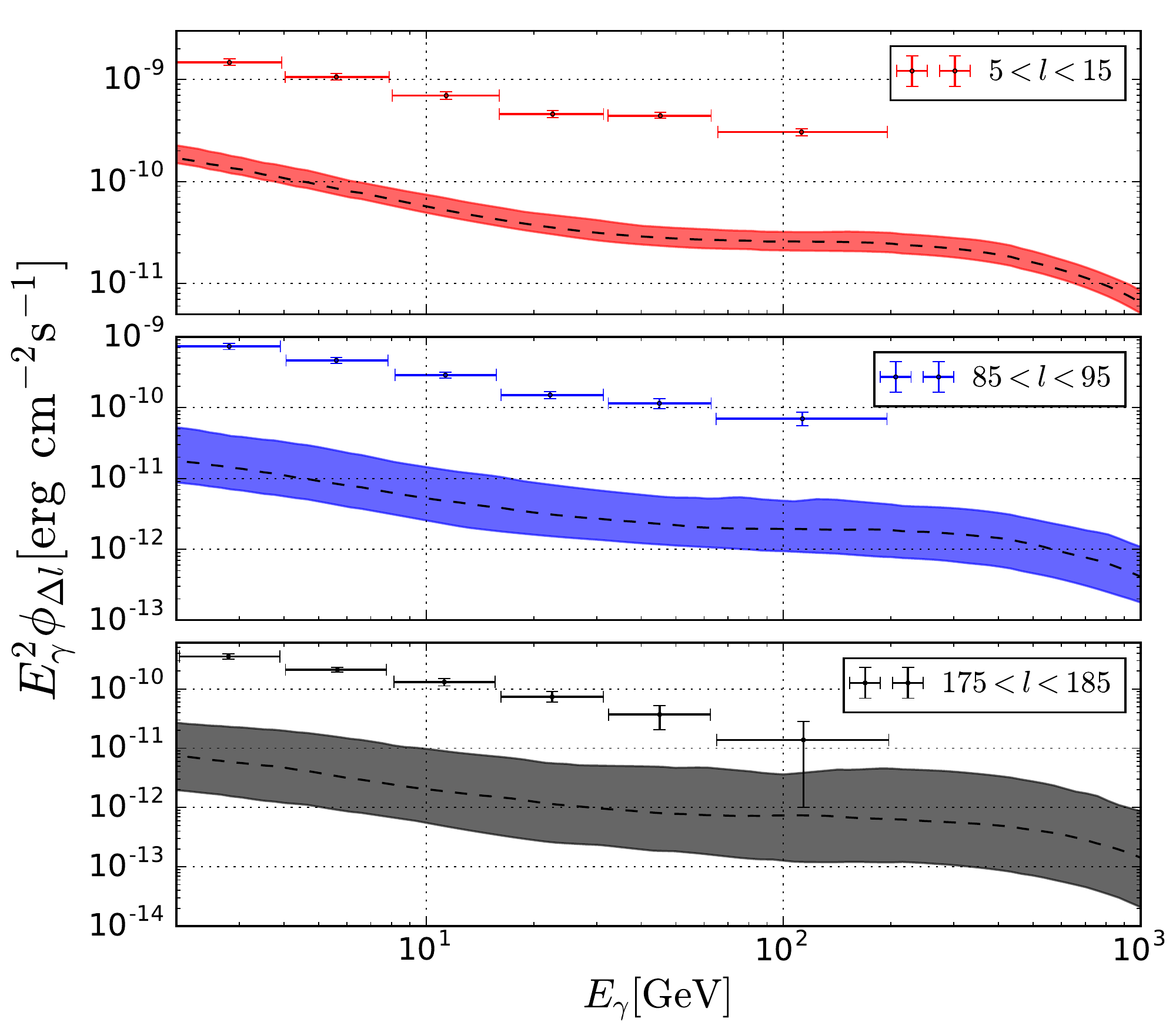} 
\caption{Spectral energy distribution of the total gamma-ray emission $E_{\gamma}^2 \times \phi_{\Delta l}$ in three different angular sectors, $5^{\circ}\le l \le 15^{\circ}$ (top), $85^{\circ}\le l \le 95^{\circ}$ (center) and $175^{\circ}\le l \le 185^{\circ}$ (bottom). The left panel shows the case (1) of fully ionized ISM with $n_i = 0.45 \,  \mathrm{cm}^{-3}$ while the right panel shows the case (2) with  $n_i = 0.45 \,  \mathrm{cm}^{-3}$ and $n_H = 0.05 \,  \mathrm{cm}^{-3}$. The black dashed line corresponds to the mean value, while the band corresponds to a confidence level of $68 \%$. The green line in the left panel corresponds to the mean diffuse gamma-ray emission, while the orange line is the sum of the average gamma-ray emission produced by CR-halos plus the mean diffuse $\gamma$-ray emission (sum of the black dashed line and the green line).
The data points are from \protect\cite{Yang2016}.}
\label{fig:diff_flux}
\end{figure}

Now, to account for the contribution to the $\gamma$-ray emission from all CR halos around sources, we sample the SNR distribution in the Galaxy approximating the disc as infinitely thin with a radius of 16.5 kpc. In such a way, each SNR is identified by the two polar coordinate $R\in [0, R_{\rm disc}]$ and $\phi\in [0, 2\pi]$. The SNR distribution is taken as uniform in the azimuthal coordinate, $\phi$, and according to that reported by \cite{Green2015} in terms of galactocentric distance:
\begin{equation}
g_{SN,R}(R) = A \left( R/R_{\odot} \right)^{a} \exp \left[ -b (R - R_{\odot}) /R_{\odot}) \right]\ ,
\label{eq:rad_dist}
\end{equation}
where $A = 1/\int_0^{R_{\rm disc}} g_{SN}(R) dR$ is a normalization constant, $a = 1.09$, $b = 3.87$ and $R_{\odot} = 8.5$ kpc.
We further assume that the time at which a SN explodes follows a Poisson distribution where the mean rate of explosion is $1/30 \,  \mathrm{yr}^{-1}$. The sources that contribute to the total $\gamma$-ray emission are those with an age that does not exceed the maximum confinement time, $t_{\max}$, which we take equal to the confinement time of 10 GeV particles in each of the considered cases.
We can now compare our results with the analysis reported by \cite{Yang2016} where the diffuse $\gamma$-ray flux inferred from the Fermi-LAT data is reported for different angular sectors. Therefore, we switch to a system of coordinates, $(d,l)$, centered on the Sun, with $d$ the distance from the Sun and $l$ the Galactic longitude. The $\gamma$-ray flux integrated in a particular angular sector $\Delta l=l_1-l_2$ results from the sum on all SNe exploded in the relevant field of view with the appropriate age, i.e.:
\begin{equation}
\phi_{\Delta l}(E_{\gamma}, \Delta l) = \sum_{t_{\rm age} \leq t_{\max}} \sum_{l_1 \leq l \leq l_2} \frac{q_{\gamma}(t_{\rm age}, E_{\gamma})}{4 \pi d^2} \, ,
\label{eq:felix}
\end{equation}
where $t_{\rm age}$ is the time passed since explosion.
In Fig.~\ref{fig:diff_flux} we compare our results obtained using an injection slope $\alpha = 4.0$, with those by \cite{Yang2016} obtained for three different sectors $\Delta l$ and integrated over Galactic latitude $-5^{\circ} < b < 5^{\circ}$.
In order to take into account the statistical fluctuation of the spatial and temporal SNR distribution, we consider 100 different realizations and for each realization we compute $\phi_{\Delta l}$. In the figures we represent the statistical uncertainty with bands that include the 68\% of the considered realizations, while the central dashed lines represent the average value of the gamma-ray flux. Notice that the width of the uncertainty band  increases moving towards outskirts of the Galaxy, as a consequence of the reduced number of SNRs.
Fig.~\ref{fig:diff_flux} clearly shows that in the case of fully ionized medium (left plot) the $\gamma$-ray emission from the halos can significantly contribute to the observed diffuse emission at all detected energies, namely up to $\sim 100$ GeV. More precisely, at 10 GeV  the contribution due to the CR-halos is $\gtrsim 50\%$ in the sector $5^{\circ}\le l \le 15^{\circ}$ and $\approx 10\%$ in the other two sectors. Clearly this result has to be considered as an upper limit, because we do not expect that all SNRs expands into a fully ionized medium. On the contrary, when neutral are present as in the case (2), showed in the right panel, the contribution of the CR-halos strongly decreases down to $few \%$ of the detected $\gamma$-ray flux.

Our predictions are also compared with the diffuse $\gamma$-ray background produced by the interaction of the CR pool with the ISM gas, shown in the left panel of Fig.~\ref{fig:diff_flux} with a solid-green line. To perform this calculation we used the Galactic CR (proton) spectrum measured by AMS-02 and given by Eq.~(\ref{eq:fg}) (assumed to be spatially constant in the whole Galaxy), while for the gas distribution of the ISM we have used the cylindrically symmetric model reported by \cite{Moskalenko2002} for both $\rm H_I$ and $\rm H_{II}$. In Fig.~\ref{fig:diff_flux} we also report, as an orange line, the sum of the diffuse gamma-ray emission plus the mean contribution from the halos. It is worth noticing that the contribution due to the mean CR distribution alone cannot adequately account for the data (both in normalization and slope, as already reported by \cite{Acero2016} and \cite{Yang2016}), especially in the sector close to the Galactic Center where the data clearly show an harder slope. On the other hand the slope of $\gamma$-ray emission coming from the CR-halos is closer to the slope inferred from the data and one is tempted to conclude that this contribution may be the dominant one. A further consequence of this scenario is that the spectrum of the diffuse $\gamma$-ray emission for $|b| > 5^{\circ}$ should be steeper and close to $E^{-2.7}$ because at high latitude the presence of sources becomes negligible and the diffuse CR spectrum dominates. Interestingly enough, such a conclusion is in agreement with the result reported by \cite{Yang2016} (see their Fig 4).

\section{Conclusion}		\label{sec:conc}
The self-excited turbulence produced by streaming instability of CRs during the escape form their sources can considerably enhance the level of scattering in the near source region. As a consequence, CRs spend more time close to the source and the production of $\gamma$-rays from hadronic interactions with the circumstellar gas increases considerably. 
In this work we showed that the integration of the $\gamma$-ray emission from many halos along a given line of sight may contribute to the diffuse emission of the Galaxy in the $\gamma$-ray band. We compared the resulting $\gamma$-ray flux with a recent analysis of the diffuse Galactic background detected by Fermi-LAT.  
We showed that the signal is strongest when neutral Hydrogen is absent in the circumstellar medium, and almost saturate the detected flux especially in the directions towards the inner Galaxy. On the contrary the signal becomes negligible when even a small fraction of neutrals is present. The reason is due to the ion-neutral friction, a process which can damp magnetic waves very efficiently, hence reducing the CR confinement time in the near-source region. Nevertheless, the above discussion ignores the potential role of molecular clouds or other dense targets for gamma ray production, that while not affecting the production of waves may enhance the production of gamma rays. Hence a non negligible contribution to the $\gamma$-ray flux could result also when neutrals are present. Such a scenario will be investigated elsewhere.

\end{document}